\documentclass{Interspeech2024}

\interspeechcameraready

\title{Towards Improving NAM-to-Speech Synthesis Intelligibility using Self-Supervised Speech Models}

\name[affiliation={1,2}]{Neil}{Shah}
\name[affiliation={2}]{Shirish}{Karande}
\name[affiliation={1}]{Vineet}{Gandhi}

\address{
  $^1$CVIT, Kohli Centre for Intelligent Systems, IIIT Hyderabad, India\\
  $^2$TCS Research, Pune, India}
\email{neilkumar.shah@tcs.com, shirish.karande@tcs.com, vgandhi@iiit.ac.in}


\newcommand{\etal}{\textit{et al}.}

\begin{document}

\maketitle

\begin{abstract}
    
We propose a novel approach to significantly improve the intelligibility in the Non-Audible Murmur (NAM)-to-speech conversion task, leveraging self-supervision and sequence-to-sequence (Seq2Seq) learning techniques. Unlike conventional methods that explicitly record ground-truth speech, our methodology relies on self-supervision and speech-to-speech synthesis to simulate ground-truth speech. Despite utilizing simulated speech, our method surpasses the current state-of-the-art (SOTA) by $29.08\%$ improvement in the Mel-Cepstral Distortion (MCD) metric. Additionally, we present error rates and demonstrate our model's proficiency to synthesize speech in novel voices of interest. Moreover, we present a methodology for augmenting the existing CSTR NAM TIMIT Plus corpus, setting a benchmark with a Word Error Rate (WER) of $42.57\%$ to gauge the intelligibility of the synthesized speech. Speech samples can be found at \url{https://nam2speech.github.io/NAM2Speech/}
\end{abstract}

\section{Introduction}
\label{sec:introduction}
Verbal communication is a highly efficient form of social interaction, primarily facilitated by the intricate coordination of various physiological processes. The expulsion of air from the lungs triggers the vibration of vocal folds, and the articulation of the tongue, cheeks, and lips further shapes this airflow into speech. However, in cases of vocal tract pathology \cite{kolbrunner2010psychogenic}, characterized by complete or partial airway obstruction, the conventional process of speech production is disrupted. Furthermore, in situations involving private communication on mobile phones or conversations in places such as healthcare facilities or quiet public spaces, individuals may prefer to refrain from engaging in normal speech. Therefore, exploring alternatives by advancing speech and signal processing is crucial, driving research in Silent Speech Interfaces (SSI).

SSI represents a unique form of spoken communication where an acoustic signal is absent; the individual articulates silently without generating sound. Techniques in SSI comprehend speech content by analyzing silent articulatory movements or vibrations resulting from airflow movement across the neck. Several SSI techniques include Lip reading \cite{sahipjohn2023robustl2s}, Ultrasound Tongue Imaging (UTI) \cite{toth2018multi}, Real-Time Magnetic Resonance Imaging (RT-MRI) \cite{otani2023speech}, Electromagnetic Articulography (EMA) \cite{horn1997reliability}, Permanent Magnet Articulography (PMA) \cite{gonzalez2017direct}, Electrophysiology \cite{kapur2018alterego}, Electrolarynx (EL) \cite{kikuchi2004development}, and Electro-Optical Palatography \cite{stone2020cross}. However, many of these techniques face challenges for everyday use due to their non-real-time nature and invasive characteristics. For instance, UTI employs ultrasound to capture tongue movement, rtMRI records the mid-sagittal plane of the upper airway through MRI, and EMA measures sensory vibrations on articulators like lips and tongue, with limitations including intense vibrator and MRI equipment noise \cite{otani2023speech}, poor lighting \cite{toth2018multi}, and extended time for speech proficiency \cite{kikuchi2004development}.

Approximately two decades ago, Nakajima \etal~\cite{nakajima2003non} pioneered a non-invasive SSI technique utilizing a specialized microphone to capture flesh-conducted NAM vibrations behind the ear. Their work demonstrated speech recognition viability based on NAM vibrations collected in the Japanese language. About ten years later, Yang \etal~\cite{yang2012noise} introduced the CSTR NAM TIMIT Plus corpus, a $40$-minute dataset featuring NAM and its corresponding whisper speech in the English language. This resulted into several efforts to translate NAM vibrations into conventional speech~\cite{shah2018effectiveness, malaviya2020mspec}. However, several limitations impede current initiatives in this domain.
\begin{itemize}
    \item All these methods explicitly record ground-truth speech \cite{malaviya2020mspec}.
    \item The intelligibility and quality of the synthesized speech remains excessively low.
    \item All these approaches predicts Mel-based features from NAM vibrations, limiting their method's ability to synthesize speech in novel voice of interest.
    \item Due to the limited size of the database, current research cannot fully leverage the capabilities of modern deep learning techniques in this field.
\end{itemize}
These challenges further compound pre-existing issues in NAM vibrations, such as the absence of a fundamental frequency \cite{nakajima2003non} and attenuations in high-frequency components \cite{toda2009voice}. In this study, we propose a novel method to synthesize speech from NAM signals using recent Self-Supervised Learning (SSL) methods. Our study makes the following contributions:
\begin{itemize}
    \item Unlike other methods, our framework functions without the explicit necessity for studio-recorded ground-truth speech.
    \item To improve intellegibility of the synthesized speech, we propose a novel data augmentation technique for simulating speech in NAM voices and introduce a Dynamic Time Warping (DTW) method to optimize alignment with its corresponding speech.
    \item We introduce a Seq2Seq learning algorithm for cross-modality learning between NAM and the ground-truth speech representations in the latent space. This enables our method to effectively clone speech content to novel voices.
\end{itemize}

\section{Related work}
Pioneering work by Nakajima \etal~\cite{nakajima2003non} devised a specialized microphone to collect NAM samples and proposed to employ traditional Hidden Markov Models (HMMs) for NAM-to-text recognition. Toda \etal~\cite{toda2005nam} initially endeavored to generate direct speech from NAM vibrations. However, they encountered challenges in estimating fundamental frequency contours, resulting in converted speech lacking prosody and intelligibility. They proposed synthesizing whispers from NAMs to address this issue. Since then, the most recent efforts \cite{shah2018effectiveness,malaviya2020mspec} focused on extracting whisper representations for NAM-to-speech conversion tasks. Subsequent efforts to \cite{toda2005nam} focused on enhancing the recording device for improved design and usability~\cite{nakajima2006methods}. Significant contributions were made by Shimizu \etal~\cite{shimizu2007acoustic,shimizu2009frequency}, who explored the frequency characteristics and sensitivities of various NAM microphone designs. Shah \etal~\cite{shah2018effectiveness} employed Generative Adversarial Networks (GANs), while Malaviya \etal~\cite{malaviya2020mspec} utilized multiple auto-encoders aligned in the latent space to derive efficient speech representations. However, existing approaches heavily rely on studio-recorded ground-truth speech \cite{malaviya2020mspec}. To the author's knowledge, no study has yet demonstrated the possibility of synthesizing speech without explicitly recording studio-quality ground-truth speech data while showing improved intelligibility or explored the feasibility of employing machine-learning-based self-supervision techniques. 

Our research is related to the broader field of speech conversion, where translation occurs between two distinct modalities~\cite{prajwal2020learning,djilali2023lip2vec,sahipjohn2023robustl2s}. A significant aspect of our study is the reliance on SSL, where information extracted from the input audio serves as the label for learning representations in subsequent iterations. We achieve the Seq2Seq translation by extracting self-supervised embeddings using HuBERT \cite{hsu2021hubert} from both the input NAMs and the simulated ground-truth speech. This work is motivated by HuBERT's ability to operate directly at the raw waveform level, preventing loss of information due to input quantization and showing superior performance compared to other methods, as demonstrated in~\cite{polyak21_interspeech,kosgi2023parrottts,rekimoto2023wesper,djilali2023lip2vec}.

\begin{figure}[t]
  \centering
  \includegraphics[width=\linewidth]{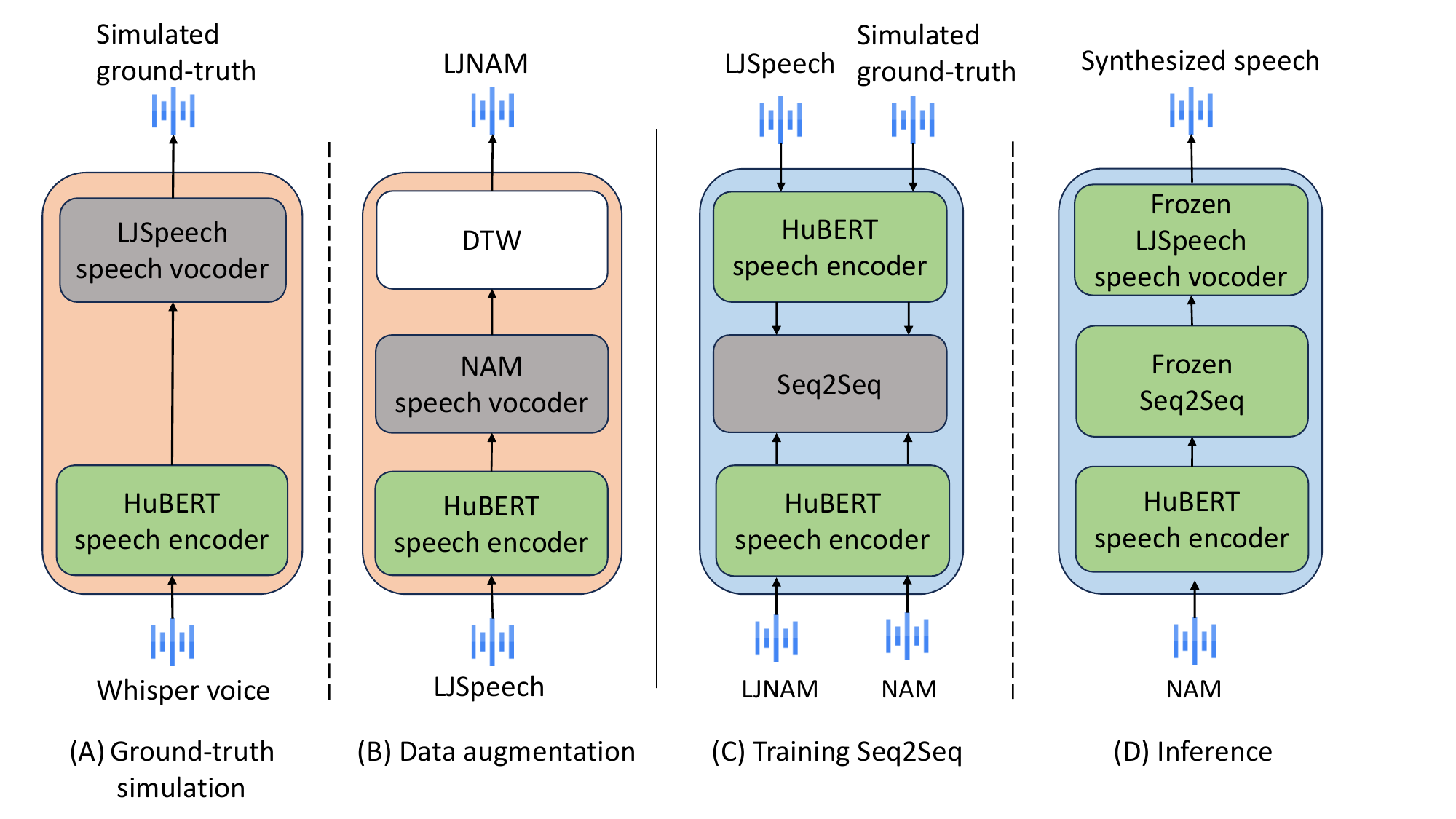}
  \caption{Proposed methodology overview: (A) Ground-truth speech simulation from whisper speech, (B) Data Augmentation with LJSpeech and DTW Algorithm to generate time-aligned LJNAM samples in a NAM-like speaking voice, (C) Seq2Seq Learning Framework, and (D) Inference Pipeline for voice Synthesis in NAM-to-Speech Conversion Task. Green boxes denote pre-trained or frozen components, while grey boxes signify training modules.}
  \label{fig:teaser}
\end{figure}

\section{Method}
Figure \ref{fig:teaser} provides an overview of our proposed method, encompassing three key stages: ground-truth speech simulation from available whisper data, data augmentation to generate additional NAM voice samples, and a Seq2Seq network trained and paired with a speech vocoder for synthesizing speech in novel voices during inference.

\subsection{Speech encoder} 
\label{sec: encoder}
Current SOTA techniques in NAM-to-speech conversion, as exemplified by \cite{malaviya2020mspec}, utilize Mel-cepstral features for encoding raw audio. However, these features encapsulate all aspects of the given audio, including speaker and ambient noise. Consequently, during training, the network is compelled to reconstruct the intended speech content in addition to the speaker and ambient noise information. This requirement complicates the training process and negatively influences the intelligibility of the converted speech and its ability to synthesize speech in novel voices.

Recent progress in widely-used SSL models like BASE HuBERT (Hsu \etal, 2021) shows promise in capturing detailed speech representations while excluding speaker and background noise information. The network employs a BERT-like masked-prediction loss on a substantial volume of unlabeled speech data. The k-means algorithm clusters the features, with cluster IDs serving as pseudo labels for a classification loss. These pseudo-labels refine the self-supervised representations iteratively. Instead of extracting cluster IDs as discrete units for the final speech representation, we choose to use $768$-dimensional embeddings for training our Seq2Seq mapping network. This approach allows us to better preserve crucial speech content during synthesis.

\subsection{Ground-truth speech simulation}
\label{sec: gt data simulation}
The current leading method in NAM-to-speech conversion \cite{malaviya2020mspec} relies on manually recording studio-quality speech data for training. Our approach departs from this conventional manual recording method and instead simulates ground-truth speech using paired whisper audio available in the CSTR NAM TIMIT Plus corpus. Figure \ref{fig:teaser}(A) illustrates our proposed simulation step. Initially, we obtain quantized HuBERT representations from utterances in the LJSpeech \cite{ljspeech17} dataset and train a speech vocoder to resynthesize speech in the LJSpeech voice. Subsequently, we acquire quantized HuBERT representations from the whisper audio and feed them into the trained vocoder to generate corresponding speech in the LJSpeech speaking style. This approach performs voice cloning from a whispering style to normal speech, which is time-aligned by the method (using one quantized unit for each 1/50th second). We rely on whisper audio, as opposed to NAM audio, to simulate ground-truth speech, driven by lower error rates and the presence of a fundamental frequency in the whisper audio.

\subsection{Data augmentation}
\label{sec: data augmentation}
The only existing corpus for NAM-to-speech conversion is limited to just $40$ minutes, posing a significant challenge for learning a transformer-based Seq2Seq framework to derive intelligible and high-quality speech representations. Recent work by Jun Rekimoto~\cite{rekimoto2023wesper} proposed the mechanical conversion of speech data to a whisper voice, to generate additional training samples, using an LPC-based audio conversion tool~\cite{zeta_chicken}. However, no such tools or techniques exist for augmenting NAM voice due to the challenges discussed in Section \ref{sec:introduction}. To address this issue, as illustrated in Figure \ref{fig:teaser}(B), we propose augmenting the existing NAM corpus using speech-to-speech synthesis techniques. Similar to section \ref{sec: gt data simulation}, we use the encoder-decoder speech cloning architecture. We first obtain quantized HuBERT representations of the NAM vibrations and train a speech vocoder to re-synthesize speech in the NAM voice. Subsequently, we derive HuBERT representations from utterances in the LJSpeech dataset \cite{ljspeech17} and pass them through the trained speech vocoder to synthesize speech in the NAM voice (speaking style). The derived HuBERT representations from the LJSpeech dataset only retain the speech content \cite{polyak21_interspeech}, neglecting the speaker and ambient noise characteristics, thus significantly benefiting our proposed data augmentation technique. This method allows us to augment the existing NAM corpus with content from the LJSpeech \cite{ljspeech17} dataset, totaling approximately 24 hours of NAM data. We refer to this augmented NAM dataset as LJNAM and its corresponding speech data as LJSpeech.

\subsection{Time alignment of representations}
\label{sec: dtw}
As discussed in section \ref{sec: gt data simulation}, the simulated ground-truth speech is time-aligned by virtue of the ground-truth data simulation technique. However, the augmented LJNAM dataset and its corresponding LJSpeech data are not inherently time-aligned. Time alignment between the input and the ground-truth is essential for learning a Non-Autoregressive (NAR) Seq2Seq architecture. To address this, we propose using FastDTW \cite{salvador2004fastdtw}, a DTW technique that aligns representations of varying lengths, allowing non-linear warping to find an optimal match. The DTW matching algorithm is applied to the extracted SSL embeddings from both signals. The algorithm starts with the assumption that the embeddings of both sequences at the $0^{th}$ frame match and then identifies the most matching frames for the shorter sequences from the available frames in the longer sequence. This method aligns the augmented LJNAM and its corresponding LJSpeech utterances, a crucial step for learning the Seq2Seq mapping between the two modalities.

\subsection{Seq2Seq network}
\label{sec: seq2seq}
We utilize a NAR transformer-based Seq2Seq network (Figure \ref{fig:teaser}(C)) to learn the mapping between the two latent spaces. Our network takes as input the SSL embeddings derived from NAM and LJNAM samples, while the SSL embeddings from simulated ground-truth speech and LJSpeech data serve as the target ground-truth representations. We opt for NAR models over autoregressive models because they generate all output tokens simultaneously in a single iteration, thus facilitating real-time applications \cite{rekimoto2023wesper}.


The encoder and decoder, each comprising six layers, consist of feed-forward transformer blocks with two multi-head self-attention mechanisms \cite{vaswani2017attention} and 1-dimensional convolutions inspired by Fastspeech2 \cite{ren2021fastspeech}. The encoder processes NAM embeddings into a sequence of fixed-dimensional vectors, while the decoder predicts ground-truth speech embeddings. For training, we set the batch size to $16$ and the maximum number of steps to $20,000$. The Adam optimizer is employed with an initial learning rate of $4.4 \times 10^{-2}$, an annealing rate of $0.3$, and annealing steps at [$3000$, $4000$, $5000$]. The HuBERT model encodes speech into embeddings at a frame rate of $50Hz$. The model utilizes Mean Squared Error (MSE) loss, quantifying the difference between the decoded and ground-truth speech embeddings. The objective can be written as:

\begin{equation}
    \mathcal{L}_{\text{MSE}} = \frac{1}{T} \sum_{i=1}^{T} ||S_{\text{ssl\_i}} - \hat{S}_{\text{ssl\_i}}||^2,
\end{equation}
where $S_{\text{ssl}}$ are the ground-truth speech embeddings, $\hat{S}_{\text{ssl}}$ are the decoded speech embeddings, and $T$ is the time-steps.

We add an extra fully connected linear layer to improve the model's capability in predicting Connectionist Temporal Classification (CTC) tokens immediately following the transformer encoder layer. The ground-truth text sequences are tokenized using the Wav2Vec2 tokenizer \cite{baevski2020wav2vec}. Let $N_{\text{ssl}}$ represent the input NAM embeddings, and $Enc_{\text{ssl}}$ denote the output of the transformer encoder. If $C$ denotes the character labels corresponding to the ground-truth text, the objective is to minimize the negative log-likelihood using $P_{CTC}(C|Enc_{\text{ssl}})$ \cite{graves2006connectionist}. It is mathematically defined as :

\begin{equation}
    \mathcal{L}_{CTC} := -\log P_{CTC}(C|Enc_{\text{ssl}}).
\end{equation}

By calculating the weighted sum of the MSE and CTC loss functions, the final objective function can be written as:

\begin{equation}
\label{eq:total-loss-ctc}
\mathcal{L}_{\text{Tot}} = \alpha_{CTC} * \mathcal{L}_{\text{CTC}} + \alpha_{MSE} * \mathcal{L}_{\text{MSE}},
\end{equation}

where $\alpha_{\text{CTC}} \in \mathbb{R}$ and $\alpha_{\text{MSE}} \in \mathbb{R}$ are the hyperparameters that balances the influence of the two loss terms. We set $\alpha_{\text{CTC}}$ and $\alpha_{\text{L1}}$ to $0.001$ and $1$, respectively.

\subsection{Speech vocoder}
\label{sec: vocoder}
The speech vocoder takes as input the speech embeddings predicted by a Seq2Seq network and synthesizes speech. We train a customized version of HiFiGAN-v2 \cite{kong2020hifi,polyak21_interspeech} for synthesizing speech from the SSL embeddings. In the generator, transposed convolutions upsample the SSL embeddings of the ground-truth speech, and a residual block is used for receptive field expansion, ultimately generating the synthesized signal. The discriminator distinguishes between the synthesized and original signals, utilizing multi-period and multi-scale networks to capture temporal patterns, details, and global structure. To enable the generation of content in the voice of the user's preference, we train a multi-speaker speech vocoder using voices available from \cite{ljspeech17,syspin}. For model configuration, we set the batch size to $16$, the learning rate to $2 \times 10^{-4}$, the number of embeddings to $100$, the embedding dimension to $128$, and the model input dimension to $256$.

\section{Dataset}
\label{cstrdataset}
We evaluate our proposed framework on the CSTR NAM TIMIT Plus corpus \cite{yang2012noise}, a publicly available dataset comprising NAM vibrations along with their corresponding whisper audio and text. The dataset comprises $421$ sentences spoken by a female speaker, extracted from the Herald text in a studio setup. The entire dataset spans $40$ minutes, with a sampling frequency of $16,000$ Hz. Consistent with prior studies \cite{malaviya2020mspec,shah2018effectiveness}, we randomly allocate $13\%$ of the data for the test set, while the remainder, along with the augmented LJNAM dataset, is used for training. Additionally, we reserve $5\%$ of the training data for computing the validation loss.

\section{Results and discussion}
In this section, we conduct a quantitative assessment of the synthesized speech, comparing it to the current SOTA method \cite{malaviya2020mspec}. We also illustrate the benefits of integrating CTC loss and data augmentation into the existing CSTR NAM TIMIT Plus corpus, to enhance the intelligibility of the synthesized speech. Unlike many existing works in this field that solely report MCD, our study goes beyond and includes error rates as a quantitative measure of intelligibility for synthesized speech. We utilize Whisper-ASR \cite{radford2023robust} to transcribe the synthesized speech, computing both WER and Character Error Rate (CER). The simulated ground-truth speech using our proposed ground-truth speech simulation (Section \ref{sec: gt data simulation}) exhibits a CER of $12.43\%$ and a WER of $24.73\%$.

\begin{table}[t]
  \caption{Recognition performance of synthesized speech using only the CSTR NAM TIMIT Plus database with no data augmentation applied.}
  \label{tab:comparison_mspecnet}
  \resizebox{0.47\textwidth}{!}{
  \begin{tabular}{ccccl}
    \toprule
    Method & CTC loss & CER $\downarrow$ & WER $\downarrow$ & MCD $\downarrow$ \\
    \midrule
    Ours & Yes & 28.90 & 48.83 & \textbf{4.716} \\
    & No & 30.45 & 49.41 & 4.754 \\
    \midrule
    DiscoGAN \cite{malaviya2020mspec} & - & - & - & 6.65 \\
    Mspec-Net \cite{malaviya2020mspec} & - & - & - & 8.19 \\
    \bottomrule
  \end{tabular}
  }
  \end{table}
  
\subsection{Recognition performance with no data augmentation}
\label{sec: res with no data aug}

Table \ref{tab:comparison_mspecnet} presents the quantitative evaluation of our proposed method without data augmentation compared to current state-of-the-art approaches. Due to the unavailability of studio-recorded ground-truth speech from the authors of MSpec-Net \cite{malaviya2020mspec}, we cannot train their model, and consequently, we cannot compute WER and CER on our test split. Their paper exclusively reports the MCD metric. The introduction of CTC loss to the Seq2Seq network resulted in improvements across all three metrics. This indicates that leveraging ground-truth text significantly enhances the learned speech representations. Remarkably, our proposed method with inclusion of CTC loss, achieves a significant decrease of $29.08\%$ and $42.42\%$ in the MCD metric compared to the current SOTA MSpec-Net and DiscoGAN frameworks. It is essential to note that MSpec-Net \cite{malaviya2020mspec} calculates the MCD metric based on studio-recorded ground-truth speech, while our method utilizes simulated ground-truth speech as discussed in Section \ref{sec: gt data simulation}.

\begin{table}[t]
  \caption{Recognition performance of synthesized speech using the CSTR NAM TIMIT Plus database with data augmentation applied.}
  \label{tab:comparison_dataaug}
  \resizebox{0.47\textwidth}{!}{
  \begin{tabular}{ccccl}
    \toprule
    Train database & CTC loss & CER $\downarrow$ & WER $\downarrow$ & MCD $\downarrow$ \\
    \midrule
    LJNAM + NAM & Yes & 25.73 & \textbf{42.57} & 4.98 \\
    LJNAM + NAM & No & 29.58 & 49.31 & 5.02 \\
    \bottomrule
  \end{tabular}
  }
  \end{table}
  
\subsection{Recognition performance with data augmentation}
\label{sec: res with data aug}
Table \ref{tab:comparison_dataaug} presents the quantitative evaluation of our proposed method when augmenting the existing corpus with LJNAM samples, derived using the method described in Section \ref{sec: data augmentation}. Although we did not observe any improvement in MCD with data augmentation compared to when no data is augmented, we did notice a significant enhancement in the intelligibility of the synthesized speech. By augmenting the existing corpus with LJNAMs and utilizing CTC loss, we achieved reductions of $12.82\%$ and $10.96\%$ in WER and CER, respectively. This suggests that augmenting the existing corpus with synthetically simulated data using our proposed data augmentation technique substantially enhances the intelligibility of the synthesized speech. Additionally, applying the data augmentation method to the LibriTTS \cite{zen2019libritts} corpus did not result in further improvement in perceived intelligibility.

\begin{figure}[t]
  \centering
  \includegraphics[width=\linewidth]{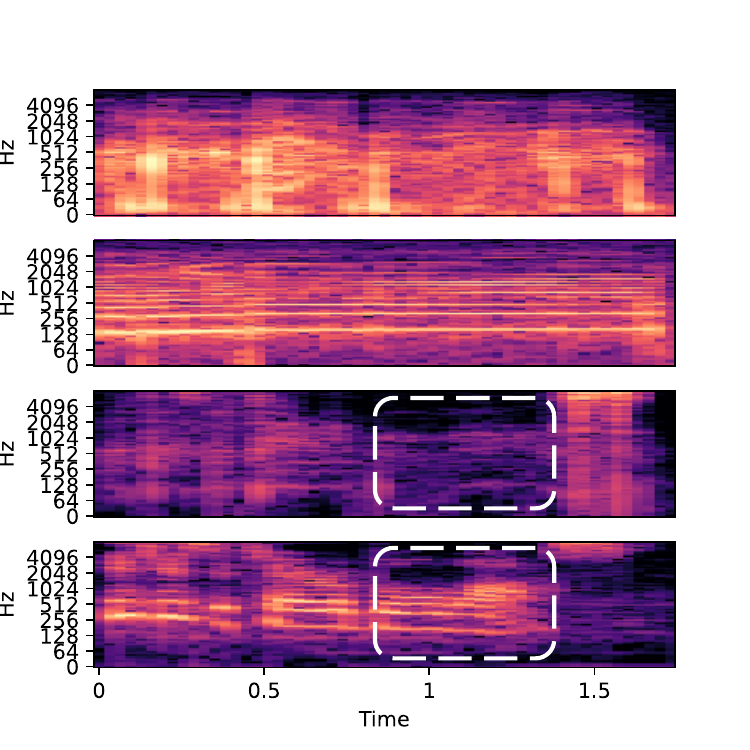}
  \caption{Mel-spectrogram comparison of (A) original NAM signal and synthesized speech from (B) DiscoGAN, (C) MSpec-Net, and (D) our proposed method. ID: 401, Text: "It is a terrible loss". The white dotted box showcases our method's superior ability to preserve and accurately estimate formants compared to MSpec-Net.}
  \label{fig:visualization}
\end{figure}

\subsection{Qualitative evaluation}
Figure \ref{fig:visualization} presents a comparison of Mel spectrograms between the original NAM signal and the synthesized speech by DiscoGAN, MSpec-Net, and our proposed method. DiscoGAN struggles to produce coherent speech, while MSpec-Net captures some intelligible content but lacks naturalness and fails to preserve lower-frequency components present in the original NAM signal. In contrast, our approach, depicted in Figure \ref{fig:visualization} (D), enhances lower-frequency formants and accurately predicts the higher-frequency formants. This dual capability significantly enhances intelligibility, establishing a novel benchmark in NAM-to-speech conversion. Our proposed setup can synthesize speech using novel voices, addressing a limitation in existing methods within this domain. Samples of these synthesized voices are available on our demo page.

\section{Conclusion}

This paper presents a novel framework for the NAM-to-speech conversion task utilizing self-supervision. Instead of relying on conventional methods to record studio-quality ground-truth speech, we employ self-supervision and speech-to-speech synthesis techniques to simulate ground-truth speech. The proposed framework, without the application of data augmentation, achieves a $29.08\%$ reduction in the MCD compared to the current SOTA method. Moreover, our data augmentation method enhances synthesized speech intelligibility, leading to a $12.82\%$ decrease in WER compared to using only samples from the existing corpus for training. Additionally, we demonstrate the model's capability to synthesize speech in novel voices. Our future aim is to enhance speech-to-speech synthesis architecture, focusing on improving the intelligibility of simulated ground-truth speech, vital for training Seq2Seq networks.


\bibliographystyle{IEEEtran}
\bibliography{mybib}

\end{document}